\documentclass{emulateapj}

\usepackage{xspace}
\usepackage{graphics,graphicx}
\usepackage{natbib}
\usepackage{multirow}
\usepackage[percent]{overpic}
\usepackage{color}

\newcommand{\as}{\mbox{\ensuremath{.\!\!^{\prime\prime}}}}

\newcommand{\asn}{$^{\prime\prime}$\xspace}

\newcommand{\nH}{$N_{\rm H}$\xspace}
\newcommand{\PL}{$\Gamma$\xspace}
\newcommand{\Msun}{$M_{\odot}$\xspace}

\newcommand{\chisq}{$\chi^2$\xspace}

\newcommand{\Chandra}{{\it Chandra}\xspace}
\newcommand{\HST}{{\it HST}\xspace}
\newcommand{\XMM}{{\it XMM-Newton}\xspace}
\newcommand{\lum}{erg s$^{-1}$\xspace}
\newcommand{\flux}{erg s$^{-1}$ cm$^{-2}$\xspace}

\shorttitle{NGC~300 X-1}
\shortauthors{Binder et al.}

\begin{document}

\title{Energy-Dependent Orbital Phases in NGC 300 X-1}
\author{B. Binder\altaffilmark{1}, J. Gross\altaffilmark{1}, B. F. Williams\altaffilmark{1}, D. Simons\altaffilmark{1}
}
\altaffiltext{1}{University of Washington, Department of Astronomy, Box 351580, Seattle, WA 98195}

\begin{abstract}
NGC~300 X-1 is a Wolf Rayet + black hole binary that exhibits periodic decreases in X-ray flux. We present two new observations of NGC~300 X-1 from the \Chandra X-ray Observatory (totaling $\sim$130 ks) along with ACS imaging data from the {\it Hubble Space Telescope}. We observe significant short-term variability in the X-ray emission that is inconsistent with an occultation by the donor star, but is consistent with structure in the outer accretion disk or the wind of the donor star. We simultaneously fit a partially-covered disk blackbody and Comptonized corona model to the eclipse egress and non-eclipsing portions of the X-ray spectrum. We find that the only model parameters that varied between the eclipse egress and non-eclipsing portions of the spectra were the partial covering fraction ($\sim$86\% during eclipse egress and $\sim$44\% during non-eclipse) and absorbing column ($\sim$12.3$\times10^{22}$ cm$^{-2}$ during eclipse egress, compared to $\sim$1.4$\times10^{22}$ cm$^{-2}$ during non-eclipse). The X-ray spectra are consistent with the movement of the X-ray source through the dense stellar winds of the companion star. From our new \HST imaging, we find the WR star within the X-ray error circle, along with additional optical sources including an AGB star and an early-type main sequence star. Finally, we use our egress measurement to rephase previous radial velocity measurements reported in the literature, and find evidence that the velocities are strongly affected by the ionization of the wind by the compact object. Thus, we argue the inferred mass of the black hole may not be reliable.

\end{abstract}

\section{Introduction}
NGC~300 X-1 (hereafter, X-1) is a high-mass X-ray binary (HMXB), located 2.0 Mpc away in the spiral galaxy NGC~300 \citep{Dalcanton+09}. The X-ray and optical emission has been studied by multiple authors, and all observations are consistent with the system consisting of a black hole (BH) + Wolf-Rayet (WR) binary \citep{Carpano+07b,Crowther+07,Barnard+08,Crowther+10,Binder+11}. The BH primary has been estimated to have a mass of 20$\pm$4 \Msun \citep{Crowther+10}, making it the second largest stellar mass BH known \citep[after another WR+BH binary IC~10 X-1,][]{Silverman+08}. Heavy stellar mass BHs are expected to form in low metallicity environments, and both NGC~300 and IC~10 have measured metallicities that are $\sim$15-30\% solar \citep{Crowther+03,Urbaneja+05}.

While numerous other BH-HMXBs with massive main sequence companions have been observed \citep[see ][ for a review]{McClintock+06}, BH + WR systems are exceedingly rare. Only one other X-ray source, IC~10 X-1, is a confirmed BH + WR binary \citep{Silverman+08,Clark+04,Bauer+04}, while Cyg X-3 in the Milky Way is a BH + WR candidate \citep{Lommen+05}. Population synthesis models suggest that $\sim$1 BH + WR binary may exist as a bright X-ray source in a Milky Way-sized galaxy \citep{vanKerkwijk+96,Lommen+05,Zdziarski+13}.

The orbital period of X-1 is short, $\sim$33 hr \citep{Carpano+07a, Crowther+10}. Periodic decreases in the X-ray flux, observed with both {\it Swift} and \XMM \citep{Carpano+07b}, have been attributed to a grazing eclipse by the donor star. This has led to a rough constraint on the inclination angle of the system through geometrical arguments (not formal dynamical modeling) of $i=60-75^{\circ}$ \citep{Crowther+10}. While similar periodic decreases in flux have been observed in other sources \citep[e.g., IC~10 X-1 and other Galactic sources; see ][and references therein]{Barnard+14}, it is not yet known whether the X-1 light curve is revealing an eclipse (a complete or partial occultation by the donor star) or a dip (where changes in the X-ray flux are due to structure in the donor star winds or an extended corona). Light curves of eclipsing sources are energy-independent, while dipping sources show periodic changes in X-ray hardness that correspond to changes in the flux level.

We have obtained two new, deep \Chandra observations of NGC~300, both of which contained X-1 in the field of view. Additionally, we have obtained optical imaging of X-1 using the {\it Hubble Space Telescope} (\HST), allowing us to study both the X-ray and optical sources in detail. In this paper we present our observations and data reduction techniques (\S\ref{section_observations}) and an analysis of the observed X-ray variability (\S\ref{section_variability}) and spectra (\S\ref{section_spectrum}). We then discuss the implications for mass accretion onto the BH and the structure of the WR winds (\S\ref{section_discussion}), and conclude with a summary of our findings (\S\ref{section_summary}).

\section{Observations and Data Reduction}\label{section_observations}

\begin{table*}[ht]
\setlength{\tabcolsep}{4pt}
\centering
\caption{\Chandra Observation Log \& Alignment to USNO-B1.0}
\begin{tabular}{cccccccc}
\hline \hline
\multirow{2}{*}{Obs. ID} & \multirow{2}{*}{Date}	& Effective Exposure	& Read 	& Data 	& \# Sources Used	& rms Residuals	& Percentage	\\
			&				& Time (ks)		& Mode	& Mode	& for Alignment		& (arcsec)			& Improvement	\\
 (1)			& (2)				& (3)      			& (4)	 	& (5) 	& (6)				& (7)				& (8)			\\
\hline
12238		& 2010 Sept. 24	& 63.0			& timed	& vfaint	& 7		& 0\as267		& 55.44\%		\\
16028		& 2014 May 16-17	& 63.9			& timed	& faint	& 5		& 0\as429		& 7.01\%		\\
16029		& 2014 Nov. 17-18	& 61.3			& timed	& faint	& 9 		& 0\as423		& 20.98\%		\\
\hline \hline
\end{tabular}
\label{table_log}
\end{table*}

\begin{table*}[ht]
\centering
\caption{Summary of \Chandra X-1 Observations}
\begin{tabular}{ccccccc}
\hline \hline
\multirow{2}{*}{Obs. ID} & R.A.	& Decl. 			& Significance		& net counts	& bkg. counts	& Off-Axis 		\\
			& (J2000)		& (J2000)			& ($\sigma$)		& (0.35-8 keV)	& (0.35-8 keV)	& Angle ($^{\circ}$)	\\
 (1)			& (2)			& (3)      			& (4)	 			& (5) 		& (6)						\\
\hline	
12238		& 00:55:10.00	& -37:42:12.2		& 177			& 2171$\pm$48	& 125$\pm$1	& 4.86	\\
16028		& 00:55:10.02	& -37:42:12.3		& 273			& 2501$\pm$51	& 66$\pm$1	& 4.85	\\
16029		& 00:55:9.99	& -37:42:12.1		& 225			& 1692$\pm$42	& 41$\pm$1	& 4.84	\\
\hline \hline
\end{tabular}
\label{table_X1_summary}
\end{table*}

	\subsection{X-ray Observations from \Chandra}
We have obtained two new, deep observations of the spiral galaxy NGC~300 using the \Chandra Advanced CCD Imaging Spectrometer (ACIS-I) taken on 2014 May 16-17 and 2014 November 17-18. X-1 was contained within the field of view of both observations. We supplement our new observations with a publicly available \Chandra ACIS-I observation of NGC~300 from 2010 \citep{Binder+11}. All X-ray observations were reduced using the \Chandra Interactive Analysis of Observations (CIAO) software package version 4.6.1 and CALDB version 4.6.3 using standard reduction procedures. The data from all three observations were reprocessed from \texttt{evt1} using \texttt{chandra\_repro}. Background light curves were extracted and no strong background flares were found. We used the task \texttt{lc\_clean} and 5$\sigma$ clipping to create good time intervals (GTIs) for each observation; all event data were filtered on the GTIs. Table~\ref{table_log} provides an observation log of all three observations, including: the observation ID number (hereafter referred to as the ``Obs. ID''), the date of the observation, the effective exposure time in kiloseconds, and both the read mode and data mode.

The \texttt{evt2} data were aligned to the USNO-B1.0 catalog using the CIAO tasks \texttt{wcs\_match} and \texttt{wcs\_update}. Exposure maps were constructed using the CIAO tool \texttt{fluximage}, which automatically produces exposure-corrected images using a user-specified instrument map. We assume spectral weights appropriate for both XRBs and AGN: a power-law spectrum (with \PL= 1.7) absorbed by the average foreground column density \citep[\nH = 4.09$\times10^{20}$ cm$^{-2}$,][]{Kalberla+05}. We additionally use the task \texttt{reproject\_image} to merge all three exposures into a single image. Point sources in each exposure and the merged image were detected using \texttt{wavdetect}, and sources with a signal-to-noise ratio $\geq3$ were considered significant.

X-1 was detected at high significance in all three observations, at positions in excellent agreement with the position reported in SIMBAD\footnote{See \url{http://simbad.u-strasbg.fr/}.}. Table~\ref{table_X1_summary} summarizes the observed properties of X-1 in each individual exposure, including the \texttt{wavdetect} R.A. and Decl., the off-axis angle from the nominal aim-point in each observation, detection significance, and net and background counts (in the 0.35-8 keV band). We use PIMMS\footnote{See \url{http://cxc.harvard.edu/toolkit/pimms.jsp}.} v4.2 to estimate the level of pile-up suffered during each observation and find the fraction to be low, $<5$\%. 

\begin{figure}[ht]
\centering
\includegraphics[width=0.8\linewidth,clip=true,trim=0cm 0cm 0cm 0cm]{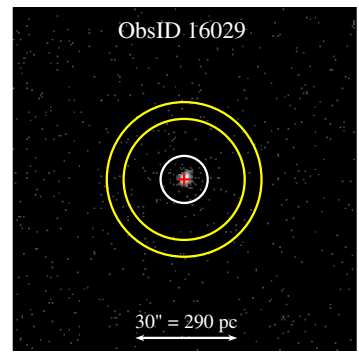} 
\caption{\Chandra 0.35-8 keV image of X-1 in ObsID 16029. The red cross indicates the position of the X-ray source, while the white circle shows our source extraction region and the yellow annulus shows the region used for background extractions.}
\label{figure_postage_stamps}
\end{figure}

\begin{figure}[ht]
\centering
\includegraphics[width=1\linewidth,clip=true,trim=0cm 0cm 0cm 0cm]{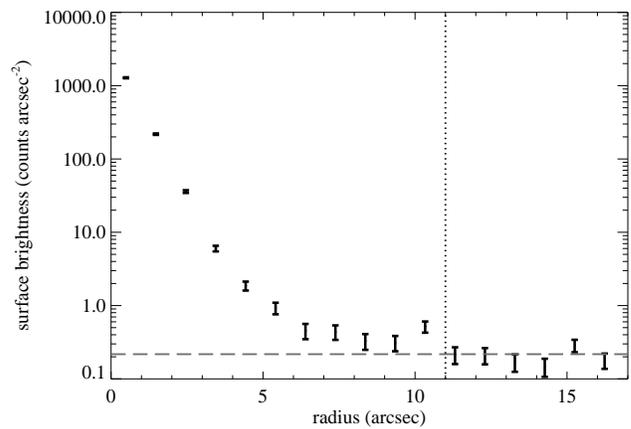} 
\caption{The radial surface brightness profile of the merged X-1 image. The dashed gray line shows the background level, and the vertical dotted line shows the inner radius of the background extraction region.}
\label{figure_rprofile}
\end{figure}

Figure~\ref{figure_postage_stamps} shows a 0.35-8 keV image of X-1 from ObsID16029 (the observation from ObsID 16028 looked similar). We find no evidence for diffuse or extended emission in the vicinity of X-1. The source region shown in white (used to extract light curves and spectra) was chosen to be a circular region, centered on the merged source coordinates, with a radius that enclosed $\sim$90\% of the \Chandra PSF at that position. To determine the background region (the circular annulus shown in yellow), we first extracted a radial surface brightness profile from the merged \Chandra image of X-1. The inner radius was set at the distance from the source where the radial surface brightness profile reached background levels, as shown in Figure~\ref{figure_rprofile}. The outer radius was then determined such that the source and background regions contained approximately the same number of pixels. Source and background regions were examined by eye to ensure that no other obvious sources were observed in either region.

	\subsection{Optical Observations from \HST}\label{HST_imaging}
In addition to our two new \Chandra exposures, we have new {\it Hubble Space Telescope} (\HST) imaging to study the resolved stellar populations in the immediate vicinity of X-1. To directly compare our X-ray observations with the optical ones, we first needed to place both sets of observations on the same coordinate system. 

Directly aligning both the \Chandra and \HST images was not reliable due to the difficulty matching X-ray sources with optical counterparts in the \HST field. We therefore performed relative astrometry by aligning both the X-ray and optical images to the same large-field optical reference image and coordinate system. To do this, we retrieved a publicly-available $B$-band image from NED\footnote{\url{http://ned.ipac.caltech.edu/}} \citep{Kim+04}. The reference $B$-band image was aligned using the USNO-B1.0 catalog using the positions of six stars. The IRAF task \texttt{ccmap} was used to compute the plate solution and to update the image header with corrected WCS information; the root-mean square (rms) residuals of the fit were 0\as158 in right ascension and 0\as159 in declination. We were next able to align our \HST fields to the ground-based $B$-band image by identifying five matched sources. The resulting rms residuals from the $B$-band/\HST alignment were 0\as125 in right ascension and 0\as0625 in declination. Adding in quadrature the rms residuals for both fields produces a final optical astrometry rms of 0\as201 in right ascension and 0\as171 in declination.

\begin{figure}[ht]
\centering
\begin{tabular}{cc}
\includegraphics[width=0.44\linewidth,clip=true,trim=4.55cm 2.75cm 4.55cm 1.45cm]{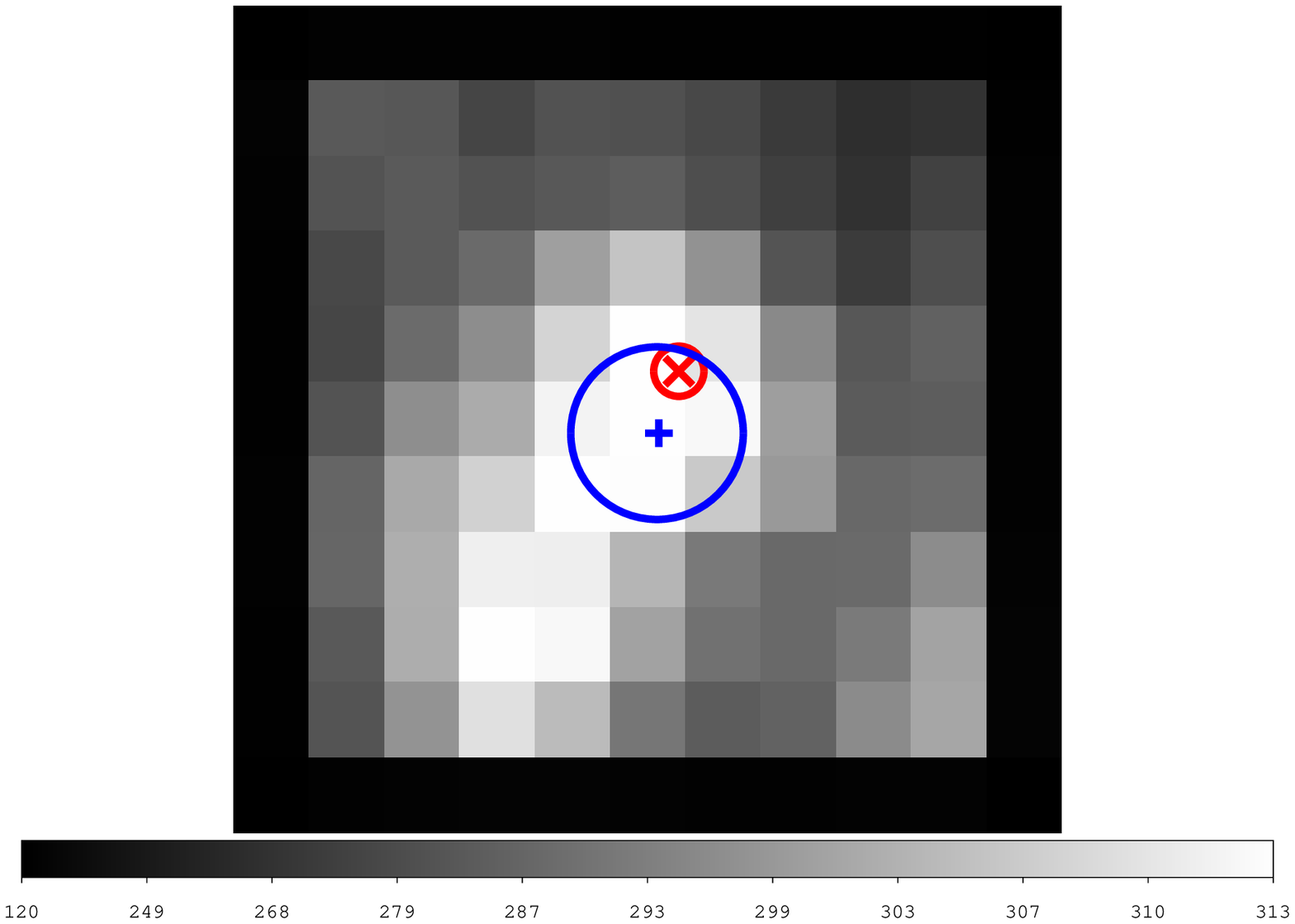} &
	\begin{overpic}[width=0.472\linewidth,clip=true,trim=0.5cm 0.5cm 0.5cm 0.5cm]{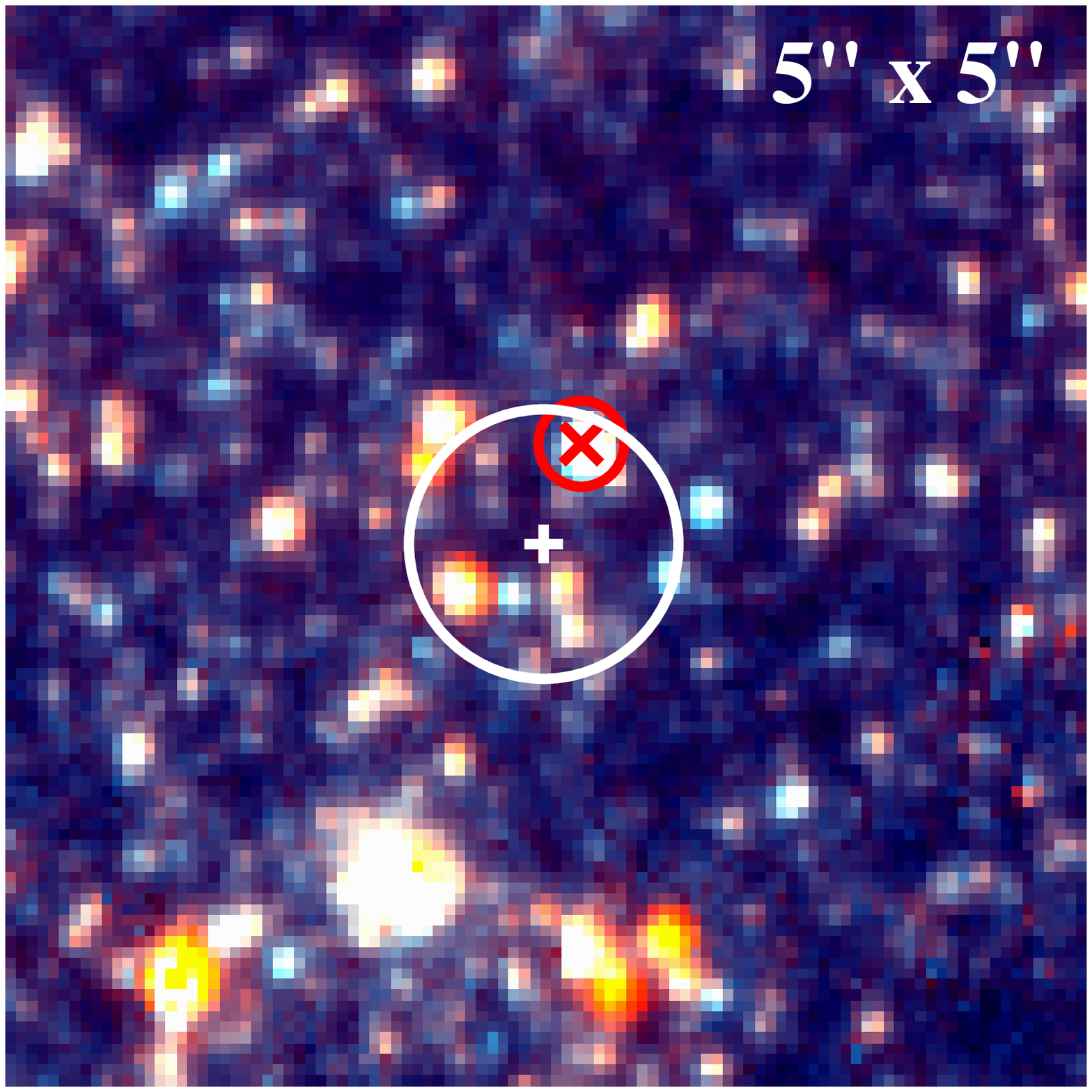}
 		\put (28,39) {\Large {\textcolor{red}{$\rightarrow$}}}
		\put (44,30) {\Large {\textcolor{cyan}{$\uparrow$}}}
	\end{overpic}	\\
\end{tabular}
\caption{{\it Left}: Smoothed, 5\asn $\times$ 5\asn ground-base $B$-band image of NGC~300 X-1. The position of the X-ray source and the corresponding X-ray error circle are shown in blue, and position of the WR star is shown in red. {\it Right}: The optical, RGB-rendered \HST image of NGC~300 X-1. The $F814W$ exposure is shown in red, $F606W$ is shown in green, and the blue band is estimated as 2$\times G - R$. The box is 5\asn on a side. The position of the X-ray source and the corresponding X-ray error circle are shown in white; the WR candidate position and optical error circle are shown in red. The red arrow indicates the AGB star, and the cyan arrow indicates the likely high-mass main sequence star.}
\label{figure_HST_aligned}
\end{figure}

We searched for optical \HST counterparts in the vicinity of the X-1. The error circle had a radius of 0\as69; fourteen optical sources were found within the X-ray error circle, including a likely WR star. The likely WR star is detected at an RA (J2000) of 00:55:9.99 and Decl. (J2000) -37:42:12.65. This position is $\sim$0\as5 off from the previously reported position of the WR from ground-based observations \citep{Schild+03,Carpano+06,Crowther+07}. However, the seeing associated with the ground-based observations was on the order of $\sim$1\asn, and the region is crowded with other relatively bright stars, making our observed position consistent with earlier observations. In Figure~\ref{figure_HST_aligned} we show the position of the X-ray source and the WR star, with their associated error circles, for both the \HST image and the ground-based $B$-band image. To create an RGB-rendered image from our \HST observations, the $F814W$ image is set to the red channel, the $F606W$ filter is used for the green channel, and the blue channel is estimated by taking $2\times F606W-F814W$.

To measure resolved stellar photometry on the \HST imaging data, we used the same techniques and software as those applied to the ACS imaging of the Panchromatic Hubble Andromeda Treasury \citep{Williams+14}.  In short, we used the point spread function photometry package DOLPHOT - an updated version of HSTPHOT \citep{Dolphin00}.  In DOLPHOT, all of the individual CCD exposures are aligned and stacked in memory to search for any significant peaks, and then each significant peak above the background level is fitted with the appropriate point spread function.  The measurements are corrected for charge transfer efficiency and calibrated to infinite aperture.  The measurements are then combined into a final measurement of the photometry of the star.  The final measurements include a combined value for all data in each band for the count rate, rate error, VEGA magnitude and error, background, $\chi$ of the PSF fit, sharpness, roundness, crowding, and signal-to-noise.  For our photometry reported here, we use the combined VEGA magnitude in each observed band.

We measure the $F606W$ and $F814W$ magnitudes of the WR candidate to be $m_{606}=22.412\pm0.005$ and $m_{814}=22.327\pm0.007$, consistent with ground-based estimates of the $V$-band apparent magnitude of 22.53 from \cite{Carpano+06} and 22.44 from \cite{Schild+03}. Using a distance of 2 Mpc to NGC~300 \citep{Dalcanton+09}, these apparent magnitudes correspond to absolute magnitudes in $F606W$ and $F814W$ of -4.13 and -4.20, respectively, assuming Galactic foreground extinction values of $A_V$=0.035 and $A_I$=0.019 from \cite{Schlafly+11}, a recalibration of the \citep{Schlegel+98} dust maps. Optical spectroscopy of this star from the VLT/FORS2 shows a WR star with an early-type WN~5 spectrum \citep{Crowther+07}, which typically have a mean intrinsic absolute $V$ magnitude of -4.2 \citep{Nugis+00}. From the spectrum, \cite{Crowther+10} estimate the surface temperature of the star to be $T_*$=65000 K, with a luminosity log$(L/L_{\odot})$=5.92 and a wind speed $v_{\infty}$=1300 km s$^{-1}$.

\begin{figure}[ht]
\centering
	\begin{overpic}[width=1\linewidth,clip=true,trim=0cm 0cm 0cm 0cm]{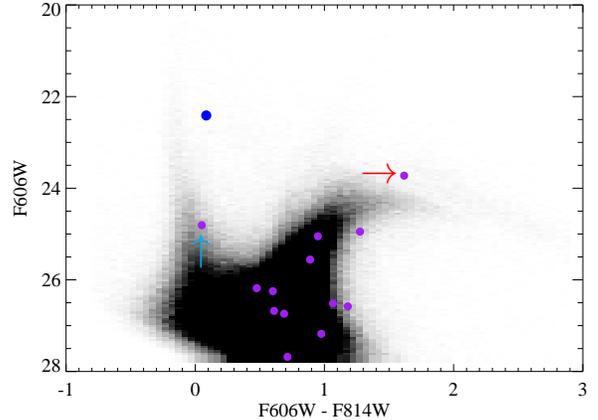}
 		\put (61,39) {\Large {\textcolor{red}{$\rightarrow$}}}
		\put (35,27) {\Large {\textcolor{cyan}{$\uparrow$}}}
	\end{overpic}	\\
\caption{A color-magnitude diagram showing all the stars within the \HST field. Purple circles indicate the optical sources found within the \Chandra error circle. The point representing the WR star is shown in blue, and is significantly brighter and bluer than the other stars within the error circle. The red arrow indicates the position of the AGB star, and the cyan arrow indicates the position of the likely high-mass main sequence star.}
\label{figure_CMD}
\end{figure}

In addition to the WR candidate, thirteen other optical sources are detected within the X-ray error circle. A color magnitude diagram (CMD) of all the sources detected within the \HST field of view is shown in Figure~\ref{figure_CMD}, with the sources falling within the X-ray error circle superimposed in purple. The WR candidate is shown in blue, and is clearly brighter and bluer than the other sources found within the error circle. Another early-type star found in the X-ray error circle is $\sim$2.5 mag fainter, and likely a main sequence star (late B-type based on an $F606W$-$F814W$ color of $\sim$0). Another very bright star (likely an AGB) is also consistent with the X-ray position. It is therefore {\it possible} that the WR candidate is not the true optical counterpart to the X-ray source. In \S\ref{section_discussion} we discuss the possibility that the optical companion to the X-ray source is a massive main sequence star or an AGB star.

X-ray-to-optical flux ratios, log$(f_X/f_V)$, can be a useful tool for discriminating between different classes of astrophysical objects; for example, in the SMC the X-ray-to-optical flux ratio combined with an optical color separates NS-HMXBs from LMXBs and AGN \citep{McGowan+08}. We therefore compute  log$(f_X/f_V)$ for {\it each} optical counterpart candidate to investigate whether {\it any} of the optical counterpart candidates are consistent with a NS-HMXB origin. We use the $F606W$ band and 2-10 keV band fluxes as $f_V$ and $f_X$, respectively. Table~\ref{table_optical_properties} lists the distance from the X-ray source position, the optical magnitudes, and log$(f_X/f_V)$ values, assuming that optical source were the true counterpart to the X-ray source. 

The NS-HMXBs in the Small Magellanic Cloud all exhibit X-ray-to-optical flux ratios $\lesssim$1 \citep{McGowan+08}. The X-ray-to-optical flux ratios computed here provide further evidence that, even if the WR star is not the true companion to the X-ray source, X-1 is inconsistent with a NS-HMXB origin. However, with the large values of log$(f_X/f_V)$ and red colors, many of the possible counterparts would be consistent with an LMXB origin. Thus, we cannot rule out the possibility of a LMXB origin from the optical photometry alone. We further discuss the nature of the optical counterpart in \S\ref{subsection_counterpart}.

\begin{table}[ht]
\setlength{\tabcolsep}{3pt}
\centering
\caption{Properties of Optical Counterpart Candidates}
\begin{tabular}{cccc}
\hline \hline
Distance from  	& \multirow{2}{*}{$m_{F606W}$} & \multirow{2}{*}{$m_{F814W}$}	& \multirow{2}{*}{log($f_X/f_{F606W}$)}	\\
X-ray source	&				&				&			\\
(1)			& (2)				& (3)				& (4)			\\
\hline
       0\as451       & 22.412$\pm$0.005       & 22.327$\pm$0.007		& 1.38 \\	
       0\as349       & 23.726$\pm$0.019       & 22.109$\pm$0.014		& 1.91 \\
       0\as108       & 24.948$\pm$0.024       & 23.673$\pm$0.015		& 2.40 \\
       0\as075       & 25.049$\pm$0.026       & 24.100$\pm$0.024		& 2.44 \\
       0\as127       & 24.808$\pm$0.022       & 24.757$\pm$0.038		& 2.34 \\
       0\as519       & 25.558$\pm$0.032       & 24.669$\pm$0.029		& 2.64 \\
       0\as053       & 26.183$\pm$0.049       & 25.707$\pm$0.062		& 2.89 \\
       0\as260       & 26.515$\pm$0.066       & 25.449$\pm$0.051		& 3.03 \\
       0\as164       & 26.580$\pm$0.089       & 25.399$\pm$0.065		& 3.05 \\
       0\as051       & 26.248$\pm$0.071       & 25.647$\pm$0.087		& 2.92 \\
       0\as049       & 26.739$\pm$0.106       & 26.051$\pm$0.118		& 3.12 \\
       0\as249       & 26.678$\pm$0.093       & 26.068$\pm$0.109		& 3.09 \\
       0\as136       & 27.178$\pm$0.156       & 26.202$\pm$0.135		& 3.29 \\ 
       0\as224       & 27.681$\pm$0.187       & 26.967$\pm$0.191		& 3.49 \\
\hline \hline
\end{tabular}
\tablecomments{The first row is the likely WR star.}
\label{table_optical_properties}
\end{table}

\section{X-ray Variability}\label{section_variability}
Variability is a common feature of many HMXBs, and the timescales over which a single source changes its X-ray luminosity can range from minutes to decades. X-1 has been previously observed to be variable, with dips in the X-ray light curve attributed to a glancing eclipse with a period of $\sim$33 hr \citep{Carpano+07b}. Geometric arguments have been used to constrain the most likely inclination of the system to be $i=60-75^{\circ}$ \citep{Carpano+07b,Crowther+10}.

\begin{figure}[ht]
\centering
\includegraphics[width=1\linewidth,clip=true,trim=0cm 0cm 0cm 0cm]{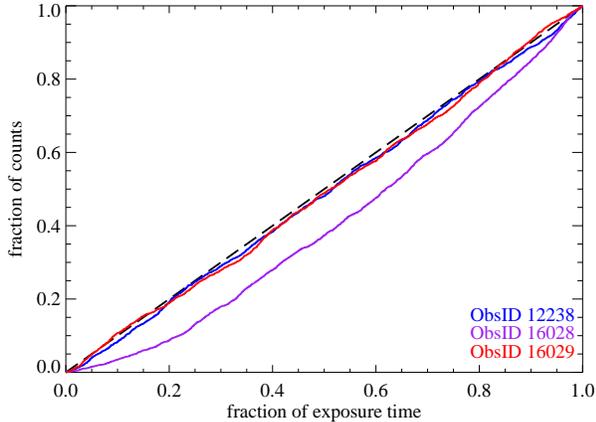}
\caption{The cumulative 0.35-8 keV photon arrival time distributions for both \Chandra observations. The dashed line shows the expected distribution for a constant count rate. The colors indicate the observation ID: 12238 is shown in blue, 16028 is shown in purple, and 16029 is shown in red.}
\label{cumulative_arrival_time}
\end{figure}

\begin{figure*}[ht]
\centering
\includegraphics[width=0.8\linewidth,clip=true,trim=0cm 0cm 0cm 0.6cm]{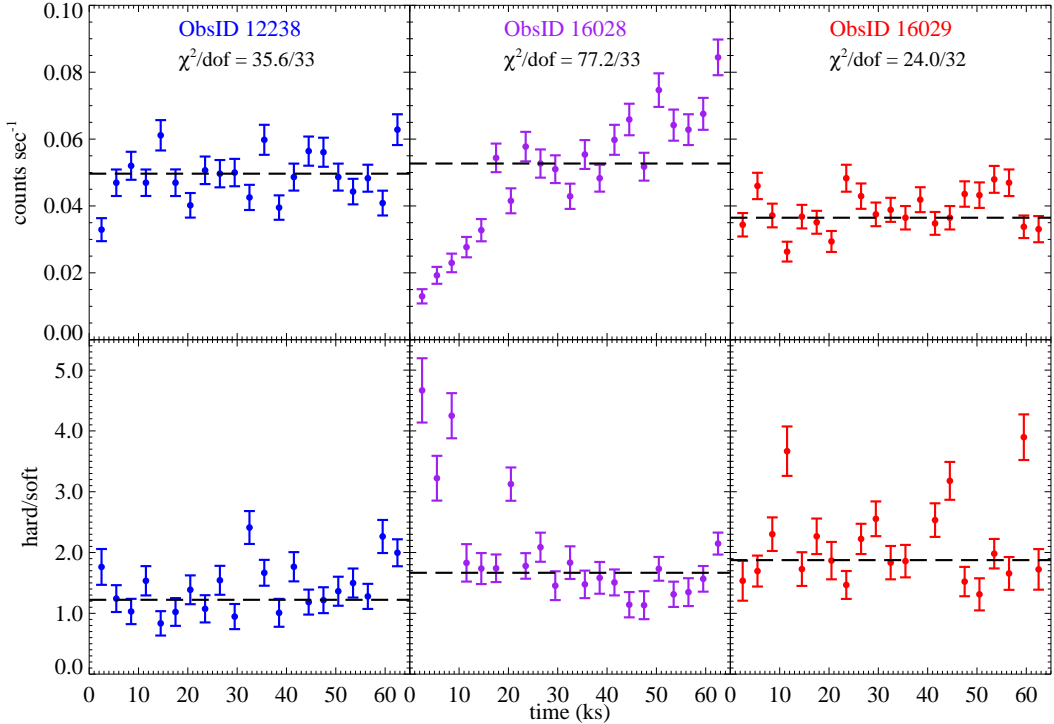}
\caption{{\it Top}: the 0.35-8 keV light curves of X-1, binned to 3 ks. The $\chi^2/dof$ value each panel shows the result from fitting to a constant count rate. {\it Bottom}: the ratio of counts observed in the hard 2-8 keV band to counts observed in the 0.35-2 keV band, also binned to 3 ks. ObsID 12238 is shown in blue (left column), 16028 is shown in purple (middle column), and 16029 is shown in red (right column). The solid horizontal line indicates the median count rates or hardness during each observation. Obs ID 16028 shows a clearly secular increase in the count rate, with a significantly harder spectrum during the first $\sim$20 ks of the observation.}
\label{figure_lc_full}
\end{figure*}

The 0.35-8 keV light curves from our \Chandra observations were extracted using the CIAO tool \texttt{dmextract}. The cumulative photon arrival time distributions are shown in Figure~\ref{cumulative_arrival_time}. The time axis is shown as a percentage of the total exposure time to account for the slight differences between the observations. The light curves, binned to 3 ks, are shown in Figure~\ref{figure_lc_full}. In ObsID 16028, X-1 showed a secular increase in 0.35-8 keV count rate. A two-sided KS test against a constant count rate (equal to the mean count rate in the observation) yielded a probability of a constant count rate of 7.5$\times10^{-3}$\%. In ObsID 16029, however, the observed light curve is consistent with a constant count rate and looks similar to the light curve observed in ObsID 12238. 

\begin{figure}[ht]
\centering
\includegraphics[width=1\linewidth,clip=true,trim=0cm 0cm 0cm 0cm]{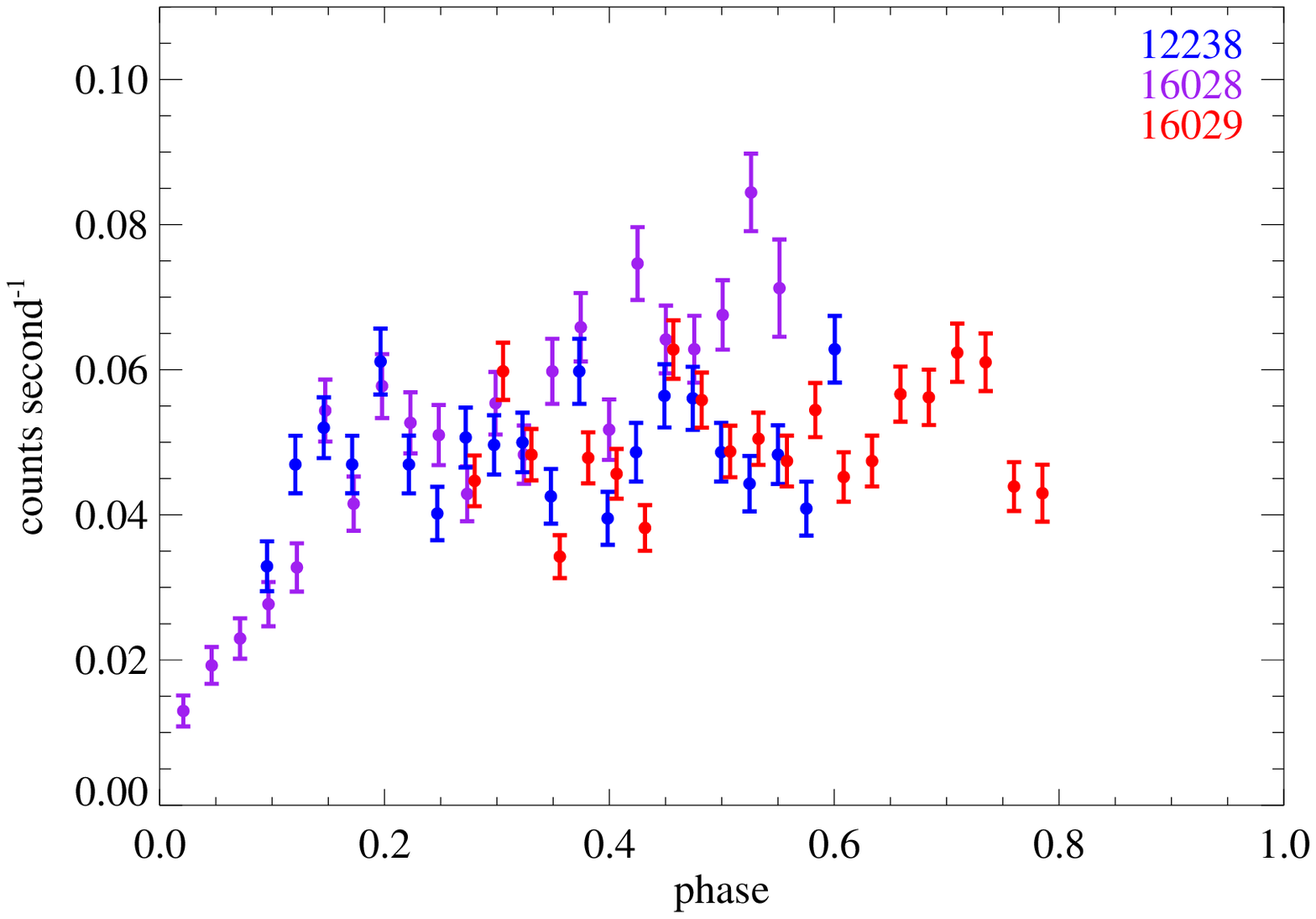}
\caption{The 0.35-8 keV light curve, folded on a period of 33.0 hr, and binned to 3 ks for clarity. This period is consistent with previous measurements from {\it Swift} and \XMM.}
\label{figure_lc_folded}
\end{figure}

\subsection{Periodicity}
The three light curves were folded over a variety of orbital periods allowed by the earlier {\it Swift} and \XMM observations (a range of 32.4-33.2 hr, in steps of 0.1 hr). The start of ObsID 16028 was set to phase zero and matches closely with the secular increase observed in \cite[][their Figure~4]{Carpano+07b}. Figure~\ref{figure_lc_folded} shows the light curve folded on a period of 33.0 hr; folding on different periods produced similar results. The first $\sim$20 ks of ObsID 12238 may represent eclipse egress, although the majority of the ObsID 12238 and 16029 exposures were taken during phases approaching conjunction.

To determine whether X-1 exhibited evidence of periodic X-ray pulsations on timescales shorter than the exposure time, we performed a periodogram analysis \citep{Horne+86} of each exposure. Monte Carlo simulations were used to calculate the 68\%, 90\%, and 99\% confidence levels assuming the light curve was dominated purely by Poisson noise. While \cite{Binder+11} reported one peak at 0.785 hours that exceeded the 90\% confidence level (but was determined unlikely to be a genuine periodic signal), we find no obvious peaks in the X-1 periodograms above the 90\% confidence level. The power density spectrum does not show significant differences from previous observations, and is well-described by a power law with a spectral index of $\sim$1. This spectral index is characteristic of accreting BHs in either the thermal or steep power law state \citep{vanderKlis94,vanderKlis95,Belloni10}. We additionally searched for periodicity in the hardness ratio light curves, but did not find any significant peaks.

\subsection{Spectral Variability}
The hardness ratio light curves for each observation are additionally shown in Figure~\ref{figure_lc_full}. If the dip in the X-1 light curve observed in ObsID 16028 was due to the donor star partially eclipsing the BH, there would be no change in the observed X-ray spectrum; one would simply observe the same spectral shape at a lower flux level. Any energy-{\it dependent} evolution in concert with the light curve would suggest the X-ray source is not being eclipsed by the companion star, but by dense stellar winds or structure in the outer accretion disk. Similar behavior has been observed in Galactic eclipsing XRBs \citep{Church+95,Barnard+01} and, more recently, IC~10 X-1 \citep{Barnard+14}. To search for energy-dependent evolution, we extracted light curves in the soft 0.35-2 keV and hard 2-8 keV energy bands. We then compared the ratio of hard/soft band counts to the total 0.35-8 keV count rate. We found no evidence of a change in hardness in ObsIDs 12238 and 16029, but ObsID 16028 exhibited an elevated hardness during the first $\sim$20 ks of the observation, corresponding to the portion of the exposure when the sharpest increase in count rate was observed. The null hypothesis that the ObsID 16028 light curve is consistent with a constant hardness ratio is rejected at the 3.6$\sigma$ level. A more detailed investigation of the spectral evolution during ObsID 16028 is discussed in the next section.

The hardness ratio peaks at the deepest point during the presumed eclipse egress and steadily decreases, reaching a relatively constant minimum level at superior conjunction with the donor star. Although our \Chandra observations do not contain the eclipse ingress, we expect that additional observations during this orbital phase would reveal an increase in hardness ratio.  At phases consistent with superior conjunction, there appears to be some low level of aperiodic variability, likely due to the BH moving through regions of the donor Wolf-Rayet stellar wind of changing optical depth; this scenario is discussed in more detail in \S\ref{section_discussion}.

\section{Spectral Modeling}\label{section_spectrum}
\begin{table*}[ht]
\centering
\caption{X-1 Spectral Models$^a$}
\begin{tabular}{cccc}
\hline \hline
Parameter		& \texttt{po} 			& \texttt{po+diskbb} 		& \texttt{po+apec}		\\
(1)				& (2)					& (3)					& (4)					\\
\hline
\multicolumn{4}{c}{ObsID 16028}	\\
\hline
\PL				& 2.39$\pm$0.07		& 2.45$\pm$0.40		& {\bf 2.29$\pm$0.12}		\\
$kT_{\rm in}$ (keV)	& ...					& 0.9$^{+0.7}_{-0.3}$	& ...						\\
$kT$ (keV)		& ...					& ...					& {\bf 1.3$^{+0.4}_{-0.3}$}		\\
\% PL			& 100				& 92					& {\bf 92}					\\
\chisq/dof			& 103/98				& 101/96				& {\bf 82/96}				\\
flux$^b$ (\flux)		& (5.8$\pm$0.1)$\times10^{-13}$ & (5.7$\pm$0.6)$\times10^{-13}$	& {\bf (5.5$\pm0.1$)$\times10^{-13}$}	\\
\texttt{ftest} probability against \texttt{po}	& ...	& 47\%				& {\bf 0.08\%}				\\
\hline
\multicolumn{4}{c}{ObsID 16029}	\\
\hline
\PL				& {\bf 2.38$_{-0.08}^{+0.09}$}	& 2.74$^{+0.67}_{-0.73}$	& 2.25$^{+0.15}_{-0.16}$		\\
$kT_{\rm in}$ (keV)	& ...					& 1.4$^{+0.6}_{-0.3}$	& ...					\\
$kT$ (keV)		& ...					& ...					& 1.1$\pm$0.3			\\
\% PL			& {\bf 100}				& 79					& 94					\\
\chisq/dof			& {\bf 66/68}			& 64/66				& 57/66				\\
flux$^b$ (\flux)		& {\bf (5.3$\pm$0.1)$\times10^{-13}$}	& (5.5$_{-0.8}^{+0.3}$)$\times10^{-13}$	& (5.0$\pm$0.2)$\times10{-13}$	\\
\texttt{ftest} probability against \texttt{po}	& ...	& 36\%				& 0.8\%				\\
\hline \hline
\end{tabular}
\tablecomments{$^a$None of our spectral fits required absorption beyond the Galactic column. We therefore kept \nH fixed at 4.09$\times10^{20}$ cm$^{-2}$ for all models. $^b$All fluxes are unabsorbed and measured in the 0.35-8 keV energy band.}
\label{table_spectralfit_results}
\end{table*}

\begin{figure*}[ht]
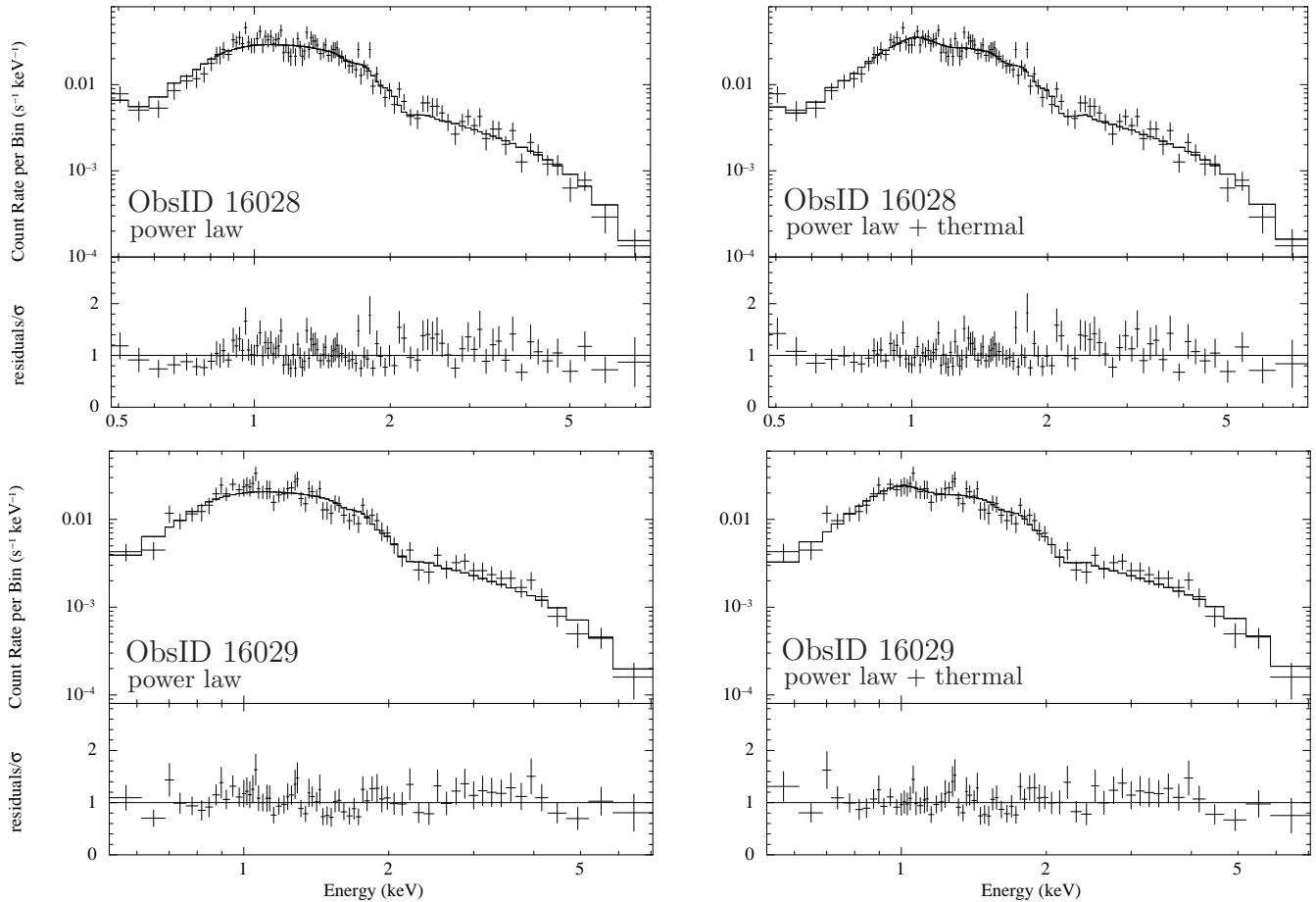

\centering
\begin{tabular}{cc}
	\begin{overpic}[width=0.33\linewidth,angle=-90,clip=true,trim=0.3cm 1.4cm 0.9cm 0cm]{16028_po_v2.eps}
 		\put (18,34) {\large ObsID 16028}
		\put (17,30) {\normalsize power law}
	\end{overpic}	& 
	\begin{overpic}[width=0.33\linewidth,angle=-90,clip=true,trim=0.3cm 2.5cm 0.9cm 0cm]{16028_po_apec_v2.eps}
 		\put (14,36) {\large ObsID 16028}
		\put (13,32) {\normalsize power law + thermal}
	\end{overpic}	\\

	\begin{overpic}[width=0.35\linewidth,angle=-90,clip=true,trim=0.3cm 1.4cm 0cm 0cm]{16029_po_v2.eps}
		\put (18,37) {\large ObsID 16029}
		\put (17,33) {\normalsize power law}
	\end{overpic}	&
	\begin{overpic}[width=0.35\linewidth,angle=-90,clip=true,trim=0.3cm 2.5cm 0cm 0cm]{16029_po_apec_v2.eps}
 		\put (14,39) {\large ObsID 16029}
		\put (13,35) {\normalsize power law + thermal}
	\end{overpic}	\\
\end{tabular}
\caption{The 0.35-8 keV spectrum of X-1 during ObsID 16028 ({\it top row}) and ObsID 16029 ({\it bottom row}). The {\it left} column shows the spectrum with a simple power law model superimposed, while the {\it right} column shows a power law plus thermal emission model.}
\label{figure_spectralfit}
\end{figure*}

We extracted spectra and response files for both new X-1 \Chandra observations in the 0.35-8 keV band using the tool \texttt{specextract} and grouped to contain at least 20 counts per energy bin. We use \texttt{XSPEC} \citep{Arnaud96} v.12.6.0q to perform all spectral fitting; the goodness of each fit is evaluated using $\chi^2$ statistics and standard weighting. All errors correspond to the 90\% confidence level, assuming symmetrical errors \citep[e.g., $\Delta\chi^2$ = 2.71, 4.61, and 6.25 for one, two, or three free parameters, respectively;][]{Lampton+76}. We assume a distance to NGC~300 of 2.00$\pm0.04$ Mpc \citep{Dalcanton+09}, and all models include neutral absorption due to the Galactic column of \nH = 4.09$\times10^{20}$ cm$^{-2}$ \citep{Kalberla+05}.

Previous studies with \XMM \citep{Carpano+07a,Barnard+08} have shown evidence that X-1 may undergo state transitions. In some observations, the X-ray spectrum is described by a power law with a photon index of $\sim$4 contributing $\sim$60\% of the total X-ray flux; the remainder of the flux originates in a multicolor disk blackbody (with $kT_{\rm in}\sim$2 keV). This is consistent with a ``thermal'' BH state \citep{McClintock+06}. However, in other observations the X-ray emission resembles a steep power law (SPL) state, with a photon index near 2.5 and an emission line near 1 keV. A longer exposure taken with \Chandra revealed a power law spectrum (\PL$\sim$2), comprising $\sim$74\% of the 0.35-8 keV flux, and a significantly cooler disk blackbody with a temperature of $\sim$0.2 keV \citep{Binder+11}. 

Since the spectrum of X-1 has previously required both power law and thermal components, we focus on three models of the 0.35-8 keV spectrum: a single power law model, a combined power law with a disk blackbody, and a power law with some emission originating in a thermal plasma (\texttt{apec}). We choose to use the \texttt{apec} component because the model, when combined with a power law, can produce excess emission near 1 keV without the need for a Gaussian component.

The results of our spectral fitting are shown in Table~\ref{table_spectralfit_results}. Although a simple power law model yielded a \chisq/dof=103/98 for the spectrum from ObsID 16028, systematic residuals were present near $\sim$1 keV. The addition of an \texttt{apec} component improved the fit residuals, with \PL=2.29$\pm$0.12 and $kT=1.3^{+0.4}_{-0.3}$ keV and \chisq/dof = 82/96. An $f$-test was used to determine if the addition of the thermal component provided a significant improvement over a simple power law fit. The addition of the thermal component produces an $f$-statistic of 0.08\%, indicating the addition of the third component improved the fit compared to the simple power law model. This model is consistent with the power law plus Gaussian emission line model utilized in \cite{Barnard+08} and \cite{Carpano+07a}, and yielded an unabsorbed 0.35-8 keV flux of ($5.5\pm0.1$)$\times10^{-13}$ \flux, equivalent to a luminosity of ($2.6\pm$0.1)$\times10^{38}$ \lum at the distance of NGC~300. Roughly 30\% of the flux of X-1 originated in the power law component.

The spectrum from ObsID 16029, however, was well described by a simple power law with \PL=2.38$^{+0.09}_{-0.08}$ (\chisq/dof = 66/68). We found no evidence that adding either a disk blackbody or a thermal component was required by the observed spectrum. The best-fit spectrum yielded an unabsorbed 0.35-8 keV flux of (5.3$\pm$0.1)$\times10^{-13}$ \flux, or (5.2$\pm$0.1)$\times10^{38}$ \lum at the distance of NGC~300.

The best-fit parameters for our various spectral models are summarized in Table~\ref{table_spectralfit_results}. Our favored spectral model is printed in boldface. Figure~\ref{figure_spectralfit} shows the 0.35-8 keV spectrum of X-1 in each observation (ObsID 16028 in the top row; ObsID 16029 in the bottom row), with either a simple power law model (left column) or a power law plus thermal emission model (right column) superimposed. The fit residuals are shown immediately below each spectrum and model. 

To test whether our current data are consistent with previous models of the X-1 spectrum, we used the best-fit models reported in \cite{Barnard+08} and \cite{Binder+11}. The fit parameters were frozen to their best-fit values, so that the only parameters allowed to vary were the respective normalizations. The resulting \chisq values are reported in Table~\ref{table_spectralfit_comparison}. ObsID 16028 was not well described by any of the previously reported best-fit models, while the ObsID 16029 spectrum showed good agreement with several earlier models. This lends further support to the notion that ObsID 16029 was observed during a similar orbital phase as Obs ID 12238 and observations 2-4 of the \XMM observations. In fact, we obtained the {\it worst} \chisq value when we modelled the ObsID 16029 spectrum with the best-fit model from observation 1 of \cite{Barnard+08}, when X-1 was observed during eclipse and egress. 

\begin{table*}[ht]
\centering
\caption{Comparing the Current Data to Past Spectral Models: \chisq/dof}
\begin{tabular}{ccccccc}
\hline \hline
			& \multicolumn{4}{c}{\cite{Barnard+08}}	&& \multirow{2}{*}{\cite{Binder+11}}	\\	\cline{2-5}
			& Obs. 1	& Obs. 2	& Obs. 3	& Obs. 4	&&				\\
			& (1)		& (2)		& (3)		& (4)		&& (5)			\\
\hline
ObsID 16028	& 139/90	& 103/91	& 114/91	& 119/91	&& 112/90		\\
ObsID 16029	& 71/68	& 67/69	& 66/69	& 69/69	&& 70/68			\\
\hline \hline
\end{tabular}
\label{table_spectralfit_comparison}
\end{table*}

	\subsection{Temporal Variations in the Spectrum of ObsID 16028}
Variations in the X-1 light curve are consistent with an orbital period of $\sim$33 hr \citep{Carpano+07b}. Although not fully captured with our \Chandra observations, the X-ray eclipse data from {\it Swift} and \XMM indicate an eclipse duration of $\sim$6 hr. Both the orbital period and eclipse duration are similar to those of IC~10 X-1, another eclipsing WR + BH binary, which additionally exhibits spectral changes during eclipsing and non-eclipsing periods \citep{Strohmayer+13,Barnard+14,Laycock+15}. The evolution of the IC~10 X-1 spectrum with orbital phase provides strong evidence that the system hosts a substantial extended corona \citep{Barnard+14}.

The sharp increase in count rate observed during the first $\sim$20 ks ($\sim$5.5 hours) of ObsID 16028 is consistent with previous observations of the X-1 eclipse egress. Furthermore, the observed energy evolution in the light curve indicates that we are not observing a simple occultation of the X-ray source by the photosphere of the companion star, and that other structures in the binary (stellar winds, an extended corona, etc.) are likely contributing to the observed periodic decrease in flux. 

To investigate the cause of the changing hardness as a function of time, we extracted two spectra from the ObsID 16028 observation (during the eclipse egress, from 0-20 ks, and from 40 ks until the end of the observation, when the X-ray source was roughly halfway through its orbit) to search for variations in the spectral shape with time. The egress spectrum was binned to contain at least 10 counts per bin, as the first 20 ks contained only $\sim$380 net counts, while the final third of the observation contained a sufficient number of counts to allow the spectrum to be binned to at least 20 counts per bin. 

\begin{figure}[ht]
\centering
\hspace{-25pt}
\includegraphics[width=0.75\linewidth,angle=-90,clip=true,trim=0cm 0cm 0cm 0cm]{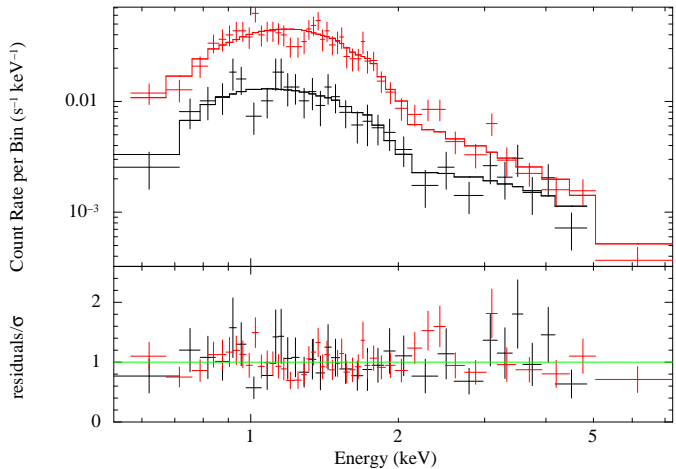}
\caption{The 0.35-8 keV eclipse egress spectrum (black) and the non-eclipsing spectrum (red). Both spectra were fit with a disk blackbody and a partially covered Comptonized component representing the extended corona. The only parameters allowed to vary between the two fits were the relative fluxes of the disk and corona, the column density of the absorber, and the covering fraction. The corona was suffered more absorption and a higher covering fraction during the eclipse egress spectrum as compared to the non-eclipse spectrum.}
\label{figure_partial_covering}
\end{figure}

We expect that the BH in X-1 hosts both an accretion disk, emitting as a disk blackbody, and a corona. We first attempted to determine whether the eclipse egress spectrum was consistent with a compact or extended corona scenario. We used the \texttt{XSPEC} models \texttt{diskbb} (to model the X-1 accretion disk) and \texttt{comptt} (to model a Comptonized corona), with both components being subject to the same absorbing column. In the \XMM observations, the best-fit spectral models yielded high inner disk temperatures ($\sim$2 keV) and power law indices \PL$>3$. In a compact corona scenario, such parameters are unphysical, as a compact corona can only access photons from the inner accretion disk \citep{Roberts+05,Goncalves+06}. In our spectral model, the compact corona is represented by forcing the inner disk temperature to be tied to the seed photon temperature. In an extended corona scenario, the corona is large enough to access soft photons from the outer edge of the disk in addition to the hot inner-disk photons \citep{Haardt+93}, therefore the disk temperature and seed photon temperature are allowed to vary independently from one another.

We first considered the eclipse egress portion of the spectrum. For the compact corona scenario, we find a best-fit inner disk temperature (and, therefore, seed photon temperature) of 0.16$\pm$0.11 keV, with 38\% of the flux originating from the compact corona. The resulting \chisq/dof was 24/25. When we allow the inner disk temperature and the seed photon temperature to vary independently (e.g., the extended corona scenario), the results do not change significantly: $kT_{\rm in}=0.20^{+0.14}_{-0.06}$ keV and the resulting \chisq/dof was 24/24. We note that we recover these fit parameters even when we initialize the fit with much hotter inner disk temperatures and lower seed photon temperatures. 

The low inner-disk temperatures derived in these models imply a strong retrograde black hole spin, which is unlikely \citep[see][and references therein]{McClintock+06}. With the relatively low number of counts present in the egress spectrum ($\sim$380), we can not statistically differentiate between the compact and extended corona scenarios using the spectral model described above. As described in the previous section, however, we do obtain thermal temperatures of $\sim$1 keV when using simpler spectral models and the full, time-averaged spectrum. We therefore expect that further observations of X-1 during the X-ray eclipse would show a preference for the extended corona scenario, as was observed in IC~10 X-1 \citep{Barnard+14,Laycock+15}.

We next compared the eclipse egress spectrum with the spectrum extracted from the final third of the ObsID 16028 observation. The X-ray eclipse may be due to the BH's motion through the dense stellar winds of the WR star, with some partial occultation by the WR star. To test this scenario, both spectra were modelled as an accretion disk (\texttt{diskbb}, initialized with an inner disk temperature near 2 keV) and an extended corona (\texttt{comptt}). Both components were subject to the same absorbing column, and we applied an additional partial covering model to the corona. As when testing for different corona types, assuming a compact corona (instead of an extended corona) did not significantly affect the quality of the resulting fits.

Both the eclipse egress and non-eclipsing spectra were simultaneously fit with this model. All parameters (except the normalizations) for the disk, corona, and absorbing column were tied between the eclipse egress and non-eclipsing spectra. The covering fraction and column density of obscuring material in our partial covering model component were allowed to vary. The best-fit model is shown in Figure~\ref{figure_partial_covering} and had \chisq/dof of 69/70. The eclipse egress spectrum is shown in black and the non-eclipsing portion of the spectrum is shown in red. The best-fit inner disk temperature $kT_{\rm in}=1.1\pm0.9$ keV and the seed photon temperature was 0.1 keV. The electron temperature was $\sim$40 keV and the optical depth was 0.17. Neither spectra showed evidence of requiring an absorbing column beyond the Galactic column, and the unabsorbed fluxes of the disk and corona did not change: the total unabsorbed 0.35-8 keV flux was $\sim9.6\times10^{-12}$ \flux, corresponding to a luminosity of 4.6$\times10^{38}$ \lum, with $\sim$90\% originating from the Comptonized corona.

The partial covering component of our spectral model exhibited variability between the two spectra. During the eclipse egress, the partial covering fraction was (86$^{+13}_{-21}$)\% with an absorbing column of $N_{\rm H,cov}$=(12.3$^{+5.4}_{-2.6})\times10^{22}$ cm$^{-2}$. During the non-eclipsing portion of the spectrum, the covering fraction dropped nearly in half (44$^{+10}_{-15}$)\%, while the absorbing column dropped by almost an order of magnitude to (1.4$^{+2.0}_{-1.0})\times10^{22}$ cm$^{-2}$. These results (summarized in Table~\ref{table_spectra_eclipse}) suggest that the observed change in flux is due to a cloud of absorbing material periodically obscuring most of the X-ray source during the binary's orbital period. We discuss the X-1 corona further in \S\ref{subsection_corona}.

\begin{table*}[ht]
\centering
\caption{Simultaneous Fitting of ObsID 16028 Eclipse Egress and Non-Eclipsing Spectra}
\begin{tabular}{ccc}
\hline \hline
Parameter		& Eclipse Egress		& Non-Eclipsing		\\
(1)				& (2)					& (3)					\\
\hline
\nH (cm$^{-2}$)	& \multicolumn{2}{c}{4.09$\times10^{20}$ (Galactic; fixed)}	\\
$kT_{\rm in}$		& \multicolumn{2}{c}{1.1$\pm$0.9 keV}				\\
photon temperature	& \multicolumn{2}{c}{0.1 keV}						\\
electron temperature & \multicolumn{2}{c}{40 keV}						\\
optical depth		& \multicolumn{2}{c}{0.17}						\\
\chisq/dof			& \multicolumn{2}{c}{69/70}						\\
unabs. 0.35-8 flux	& \multicolumn{2}{c}{9.6$\times10^{-12}$ \flux}		\\
unabs. 0.35-8 luminosity	& \multicolumn{2}{c}{4.6$\times10^{38}$ \lum}	\\
partial covering \%	& 86$^{+13}_{-21}$		& 44$^{+10}_{-15}$		\\
$N_{\rm H,cov}$	& 12.3$^{+5.4}_{-2.6} \times10^{22}$	& 1.4$^{+2.0}_{-1.0} \times10^{22}$	\\
\hline \hline
\end{tabular}
\label{table_spectra_eclipse}
\end{table*}

\section{Discussion}\label{section_discussion}

	\subsection{An Extended Corona?}\label{subsection_corona}
Our spectral modeling provides preliminary evidence that X-1 may host an extended, hot corona. The temperature of the inner edge of the accretion disk, column density of obscuring material, and inferred luminosities are all remarkably similar to those observed for IC~10 X-1 \citep{Barnard+14}. While our \Chandra observations of X-1 are not deep enough to definitely rule out the compact corona scenario, the many similarities to IC~10 X-1 (which can be observed in much greater detail due to its closer distance) suggest that the X-ray spectrum of X-1 is being produced in a similar fashion.

Observations by \Chandra and \XMM of IC~10 X-1 and other dipping XRBs often reveal asymmetric eclipses, with the eclipse ingress being steeper (spanning a shorter period of time) than the egress. While our \Chandra observations do not span a full orbital cycle, we can estimate the duration of the eclipse ingress and egress from the \XMM and {\it Swift} observations of X-1 presented in \cite{Carpano+07b}: the deepest portion of the eclipse spans $\sim$0.2 in orbital phase ($\sim$6.6 hours), while the ingress and egress occur over a phase period of $\sim$0.25 ($\sim$8 hours) and $\sim$0.15 ($\sim$5 hours), respectively. Unlike other dipping sources, the NGC~300 X-1 ingress is longer than the egress.

Using the approach of \cite{Church+04}, we estimate the size of the extended corona from the eclipse ingress using the estimation $2\pi r_{\rm D} \Delta t = 2 r_{\rm C} P$ \citep[e.g., equation~1,][]{Barnard+14}. In this equation, $P$ is the orbital period, $\Delta t$ is the eclipse ingress time, $r_{\rm D}$ is the radius of the accretion disk, and $r_{\rm C}$ is the radius of the corona. \cite{Church+04} assume the disk radius $r_{\rm D}$ to be 80\% of the Roche lobe equivalent radius $r_{\rm L1}$ \citep[see also][]{Armitage+96,Frank+02}. We estimate $r_{\rm L1}$ for X-1 using the expression derived by \cite{Eggleton83}:

\begin{equation} \label{equation_RL_equivalent}
r_{L1} = \frac{0.49a q^{2/3}} {0.6 q^{2/3} + \text{ln}\left(1 + q^{1/3} \right)},
\end{equation}

\noindent where $q = M_{\rm BH}/M_{\rm WR}$ and $a$ is the binary separation. $M_{\rm BH}$ is the mass of the black hole (20$\pm$4 \Msun), and $M_{\rm WR}$ is the mass of the WR donor \citep[26$^{+7}_{-5}$ \Msun,][]{Crowther+10}. Using this equation, we estimate $r_{\rm L1}$ to be 0.3-0.4$a$. This is consistent with that predicted by \cite{Crowther+10}, who find the Roche lobe radius to be $\sim$0.4$a$. Although the radius of the inner boundary of the WR star (e.g., where the optical depth $\tau\sim10$) is typically $\sim$1-2 $R_{\odot}$ \citep{Laycock+15}, the large outflowing winds can extend the ``photosphere'' out to $\sim$8 $R_{\odot}$ \citep[see the review by][and references therein]{Crowther07}, comparable to our estimated $r_{\rm L1}$. The outer portion of the WR ``photosphere'' is therefore capable of fueling the BH accretion disk via Roche lobe overflow.

From this estimate of the Roche lobe equivalent radius, we can now estimate the radius of the corona. Using a period $P$ of 33 hours, an ingress time $\Delta t$ of 8 hours, and an accretion disk radius of $r_{\rm D} = 0.8 r_{\rm L1} = 0.35a$, we estimate the radius of the corona to be $\sim$0.2$a$. This value is consistent with corona radii of other eclipsing sources found by \cite{Church+04}, and provides additional circumstantial evidence (from the X-1 light curve alone, and not from our spectral modeling) that X-1 hosts an extended corona. 

	\subsection{Mass Loss from Stellar Winds and Accretion onto the Black Hole}\label{subsection_counterpart}
We may shed additional light on the donor star by comparing the accretion rate of the BH to the stellar mass loss rates. Using the unabsorbed, 0.35-8 keV luminosity predicted from our simultaneous eclipse egress/non-eclipsing spectral modeling ($\sim$4.6$\times10^{38}$ \lum), we estimate that the X-1 BH (with an inferred mass of 20$\pm$4 \Msun) is accreting at $\sim$18\% Eddington (with a lower BH mass radiating at a higher fraction of the Eddington limit; see \S\ref{subsection_wind_structure}). A lower BH would be required to explained the observed X-ray emission if X-1 possessed a mild degree of beaming \citep[given the system inclination of $i\sim60-75^{\circ}$;][]{King+01}. For example, a BH mass of $\sim$10 \Msun would be required to explain the observed X-ray luminosity with a beaming factor $b\sim$0.5.

The mass accretion rate $\dot{M}$, which can be estimated from $L_X = \eta \dot{M} c^2$ ($\eta$, the radiation efficiency, is assumed to be 0.1), is $\sim$8$\times10^{-8}$ \Msun yr$^{-1}$. Since our spectral modeling revealed no significant changes in unabsorbed flux, we can assume that there are no significant changes in the mass accretion rate, and that the observed change in flux is due almost entirely to the presence of high optical depth obscuring material during the eclipse.

We can additionally estimate the wind mass-loss rate of the WR star, $\dot{m}_{\rm WR}$, using the approach of \cite{Belczynski+13}:

\begin{equation}
\dot{m}_{\rm WR} = 10^{-13} \left(\frac{L}{L_{\odot}} \right)^{1.5} M_{\odot} \text{yr}^{-1}.
\end{equation} 

\cite{Crowther+10} measure log$(L/L_{\odot})$ = 5.92 for the WR star; inputting this value into the above equation yields a wind mass loss rate of $7.6\times10^{-5}$ \Msun yr$^{-1}$. This wind mass loss rate is based on more detailed stellar evolution models than were used in \cite{Crowther+10}, with a comprehensive treatment of WR winds. As a result, our estimated $\dot{m}_{\rm WR}$ is roughly a factor of ten higher than predicted by \cite{Crowther+10} and is nearly a thousand times larger than the BH mass accretion rate $\dot{M}$. Thus, the WR winds are easily sufficient to fuel the BH accretion disk.

If the donor star were {\it not} the WR star, but instead one of the other stars observed within the X-ray error circle of our \HST imaging, the above equation may not be valid. We therefore additionally estimate the wind mass loss rates for an AGB star and B-type main sequence star (both of which were found in the vicinity of the X-ray source in \S\ref{section_observations}). 

The magnitudes of the bright, red star detected within the X-ray error circle in our \HST imaging are consistent with those of an AGB star. Mass loss during the super-wind phase of thermally pulsating AGB stars (TP-AGB) can reach up to 10$^{-5}-10^{-4}$ \Msun yr$^{-1}$ \citep[see][and references therein]{Rosenfield+14}, comparable to $\dot{m}_{\rm WR}$ calculated above, with wind speeds on the order of a few tens of km s${-1}$ \citep{Mattsson+10}. Thus, an AGB wind could power the observed X-ray emission. However, if X-1 did host an AGB donor, it would additionally imply a significantly lower BH mass. 

The mass loss rates for several OB main sequence stars in the {\it Spitzer} Cygnus X Legacy Survey \citep{Hora+08} were estimated to be a few $10^{-6}$ \Msun yr$^{-1}$ \citep{Kobulnicky+10}. Although significantly lower than our estimated WR wind rate, only $\sim$8\% of the OB stellar wind would be required to fuel the X-1 BH. However, the orbital period of X-1 makes this scenario unlikely. Using Equation~\ref{equation_RL_equivalent}, we can estimate $r_{\rm L1}$ of the donor if the masses of the binary components are known. The X-1 mass function \citep{Crowther+10} and an assumed donor mass of $\sim$10 \Msun (for typical main sequence B star) implies a lower black hole mass of $\sim$6.5 \Msun. These two masses combined with the observed orbital period imply a binary separation of $\sim13R_{\odot}$. With this orbital separation, it is unlikely the main sequence star would fill its Roche lobe, and one would expect the BH to be wind-fed and not accreting via Roche lobe overflow. Such wind-fed systems are typically two orders of magnitude fainter than X-1 \citep{Liu+05, Liu+06}.

	\subsection{The Origin of the  \ion{He}{2} $\lambda$4686 Emission Feature}\label{subsection_wind_structure}

If X-1 is truly a BH accreting via Roche lobe overflow from a WR donor, the orbital separation of the binary is small enough that the WR star would be very close to filling its Roche surface. Focused stellar wind models of Robe lobe-filling donor stars \citep{Friend+82, Gies+86} predict that the stellar winds will be highly asymmetric, with the greatest mass loss occurring in the direction of the BH. One of the predictions of this model is that the \ion{He}{2} $\lambda$4686 emission feature will be present in the optical spectrum and display radial velocity variations. While it was originally thought that such an emission feature could only be produced in the densest portions of very massive blue supergiant winds (and not the stellar photosphere), this feature may also originate near the inner Lagrangian point where the stellar winds are most dense \citep{Walborn71, Conti+74, Klein+78}. 

\begin{figure}[ht]
\centering
\includegraphics[width=1\linewidth,clip=true,trim=0cm 0cm 0cm 0cm]{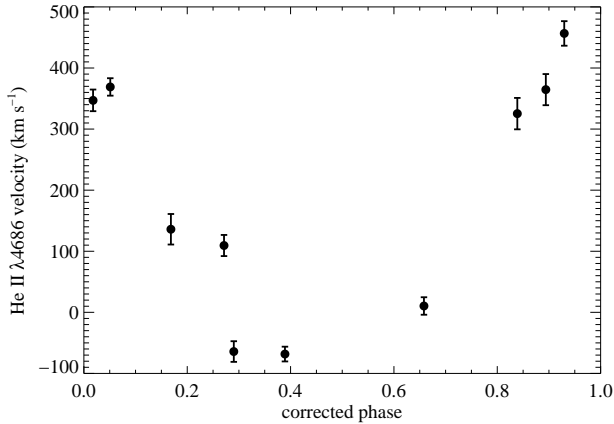}
\caption{The \ion{He}{2} $\lambda$4686 velocity measured by \cite{Crowther+10}, with phase corrected such that phase 0 corresponds to the start of our \Chandra Obs 16028 observation and the assumed period is 33.0 hours.}
\label{figure_HeII}
\end{figure}

The \ion{He}{2} $\lambda$4686 emission feature was observed in the optical spectrum of the X-1 WR donor \citep{Crowther+10}, along with radial velocity variations that are qualitatively similar to those predicted by a focused wind through a Roche lobe overflowing donor star. We use the information provided in \cite{Crowther+10} to recalculate the phase of each observation, assuming an orbital period of 33.0 hours with phase 0 corresponding to the start of our \Chandra Obs ID 16028 observation. Figure~\ref{figure_HeII} shows our results: high radial velocities correspond to the deepest part of the X-ray eclipse, with a minimum velocity occurring near the location of superior conjunction. Thus, the velocities are tracing either the excited wind in the vicinity of the BH or winds from the BH accretion disk, and not the motion of the WR star. This finding suggests that the BH mass measurement, which assumes the  \ion{He}{2} $\lambda$4686 emission line is providing the velocity of the WR star, may not be reliable. This is somewhat different from IC~10 X-1, where a phase shift of $\pi$/2 rad between the \ion{He}{2} $\lambda$4686 emission line and the X-ray light curve has been interpreted as originating in a persistent spiral shock structure in the winds of the WR donor \citep{Laycock+15}.

Our phasing of the X-ray light curve with the \ion{He}{2} $\lambda$4686 velocities suggest that, as the BH moves to opposition, it is moving through a progressively higher column density of the stellar wind. The dense portions of the wind primarily obscure the soft X-rays being emitted by the corona (which, due to its extended nature, has access to the cool outer portions of the accretion disk); a grazing eclipse by the WR donor may or may not also be occurring. As the BH moves from opposition to conjunction, the column density of obscuring material decreases and less of the corona and accretion disk are covered from the perspective of the observer. A similar toy model of how a focused wind from a Roche lobe overflowing donor can create such a change in column density was presented for the BH-HMXB Cyg~X-1 \citep[see Figure~8 of][and references therein]{Grinberg+15}, which contains an early supergiant companion \citep{Walborn73} in an even tighter binary orbit than the case of X-1 \citep[$\sim$5.5 days,][]{Gies+08} and at a lower inclination angle \citep[$i\sim27^{\circ}$;][]{Orosz+11}. The observation that the X-1 \ion{He}{2} $\lambda$4686 radial velocity shows maxima and minima at phases where the BH is expected to have the smallest radial velocities further supports this picture: at opposition, the dense winds are moving maximally away from the observer, and at conjunction the winds are moving maximally towards the observer.

This model is, of course, overly simplistic. First, the above description does not include the effects of clumps in the WR wind, which may significantly contribute to the X-ray variability. The close proximity of the X-ray source will additionally alter the wind structure dramatically, as a large portion of the wind is expected to be highly ionized. The wind velocity is additionally affected by ionization, as the winds of WR and other massive stars are line-driven \citep{Castor+75}, which in turn affects the wind density \citep{Hatchett+77}. An asymmetric wind was inferred for Cyg X-3 from radial velocity measurements of the \ion{He}{1} 2.0587$\mu$m absorption line \citep{Hanson+00}; however, the \ion{He}{1} line obtained in the X-1 optical spectrum was too weak for a similarly detailed analysis \citep{Crowther+10}. Detailed modeling of the WR wind structure is beyond the scope of this work, and the limited X-ray observations available for X-1 make it unlikely that we would be successful in distinguishing between different models. 

It was noted by \cite{Crowther+10} that the \ion{He}{2} $\lambda$4686 equivalent width ($W_{\lambda}\sim56\AA$) is a factor of $\sim$2 lower than for similar WR stars in the Milky Way. This low value of the equivalent width is possibly due to dilution from other (unresolved) stars along the line of sight. The cool giant star within the X-ray error circle may be the primary contributor to this dilution, as such stars are not hot enough to produce \ion{He}{2} $\lambda$4686 spectral features.

\section{Summary}\label{section_summary}
We have presented new \Chandra and \HST observations of the WR + BH binary NGC~300 X-1. The periodic dips in the X-ray light curves has led previous authors to conclude that the X-ray source experiences a grazing eclipse by the WR donor. We capture a large portion of an eclipse egress in one of our new \Chandra observations, and find that intervals of low X-ray flux correspond to a harder underlying spectrum. Simultaneous spectral fitting of the eclipse egress spectrum and a portion of the spectrum corresponding to superior conjunction of the BH are consistent with a partially covered multicolor accretion disk and an extended, Comptonized corona. During the eclipse egress, X-1 is partially obscured, possibly by a cloud in the donor wind, with a neutral hydrogen equivalent column density of $\sim$10$^{23}$ cm$^{-2}$ and a partial covering fraction of $\sim$86\%; however, this absorbing column density decreases by roughly an order of magnitude and the covering fraction is reduced by half by the time the X-ray source has reached superior conjunction. We argue that the observed change in the column density and covering fraction with orbital phase is due to the BH moving through dense clouds in the wind of the donor star. 

Our \HST observations reveal many optical sources within the X-ray error circle, including the WR star, an AGB star, and what is likely a high mass main sequence star. We consider the possibility that the donor star in the X-1 system is an AGB star or high mass main sequence star. While we can rule out the possibility of a main sequence donor, we cannot rule out the possibility that the AGB star is the true donor to the BH. Our observations suggest that NGC~300 X-1 is either a high-mass WR + BH binary or a low-mass AGB + BH binary, likely with an extended corona, where the observed changes in X-ray flux are largely due to the BH's motion through the dense, clumpy stellar wind of the donor star.

The \ion{He}{2} $\lambda$4686 emission feature and radial velocity variations observed in the optical spectrum of the WR star \citep{Crowther+10} are similar to those predicted by focused wind models. When we rephase the \ion{He}{2} $\lambda$4686 radial velocity measurements to match our folded X-ray light curve, we find the highest radial velocities correspond to the deepest part of the X-ray eclipse, with minimum velocities occurring near the location of superior conjunction. This is consistent with the emission feature originated from the dense material flowing through the inner Lagrangian point or originating from the BH accretion disk, and not in the low-density photosphere of the WR star.  This finding suggests that earlier BH mass measurements, which were inferred from the \ion{He}{2} $\lambda$4686 emission, may not be reliable.

\acknowledgements
The authors would like to thank the anonymous referee for suggestions that improved this manuscript, and S. Laycock for useful discussions. Support for this work was provided by the National Aeronautics and Space Administration through \Chandra Award Number G04-15088X issued by the \Chandra X-ray Observatory Center, which is operated by the Smithsonian Astrophysical Observatory for and on behalf of the National Aeronautics and Space Administration under contract NAS8-03060. This research has made use of the NASA/IPAC Extragalactic Database (NED) which is operated by the Jet Propulsion Laboratory, California Institute of Technology, under contract with the National Aeronautics and Space Administration.

\bibliography{apjmnemonic,ms}
\bibliographystyle{apj}
                                                                                             
\end{document}